\documentstyle[prb,aps,psfig]{revtex}
\def\be{\begin{equation}}
\def\ee{\end{equation}}
\def\beqa{\begin{eqnarray}}
\def\eeqa{\end{eqnarray}}
\def\ww{\tilde{w}}
\def\np{\newpage}
\def\nn{\noindent}
\def\nm{\nonumber}
\def\p{{\partial}}
\def\ll{\left}
\def\rr{\right}
\def\ra{\rightarrow}
\def\HH{Heisenberg \,}
\def\nlsm{$NL\sigma M$}

\def\b{{\beta}}
\def\g{{\gamma}}
\def\d{{\delta}}
\def\e{\epsilon}
\def\L{\Lambda}
\def\r{\rho}
\def\f{\phi}
\def\z{\zeta}

\def\om{{\omega}}

\def\H{\hat H}
\def\S{\hat S}
\def\U{\hat U}
\def\V{\hat V}
\def\rrr {\hat \rho}
\def\Q {{\bf g}}

\def \bk {{\bf k}}       
\def \br {{\bf r}}        
   
\def \bg {{\bf g}}
\def \bR {{\bf R}}
\begin{document}  
\draft
\title{\bf{The phase diagram of quantum systems: Heisenberg antiferromagnets}}
\author{
Pietro Gianinetti$^{1,2}$
and Alberto Parola$^{1,3}$
}
\address{
$^1$ Istituto Nazionale per la Fisica della Materia \\
$^{2}$ Dipartimento di Fisica, Universit\'a di Milano, Via Celoria 16,
Milano, Italy \\
$^3$ Dipartimento di Scienze Fisiche, Universit\'a dell'Insubria, Via Lucini 3
Como, Italy }
\maketitle
\begin{abstract}
A novel approach for studying phase transitions in 
systems with quantum degrees of freedom is discussed.
Starting from the microscopic hamiltonian of a quantum model,
we first derive a set of exact differential equations for the 
free energy and the correlation functions describing the 
effects of fluctuations on the thermodynamics of the system.
These equations reproduce the full renormalization group structure
in the neighborhood of a critical point keeping, at the same time,
full information on the non universal properties of the model.
As a concrete application we investigate the phase diagram of
a Heisenberg antiferromagnet in a staggered external magnetic field.
At long wavelengths the known relationship to the 
Quantum Non Linear $\sigma$ Model naturally emerges from our approach.
By representing the two point function in an approximate
analytical form, we obtain a closed partial differential 
equation which is then solved numerically. The results in
three dimensions are in good agreement with available
Quantum Monte Carlo simulations and series expansions.
More refined approximations to the general framework
presented here and few applications to other models 
are briefly discussed.
\end{abstract}

\section{Introduction}

\nn In recent years several works have appeared in the literature on the 
quantum \HH an\-ti\-fer\-ro\-mag\-net, with particular attention to the
two dimensional case, because of its deep connections to the 
physics of high temperature superconductors.  Although the corresponding 
classical model was studied by means of several techniques ranging
from high temperature expansions \cite{domb} to Monte Carlo (MC) simulations 
\cite{binder} and Renormalization Group (RG) \cite{zinn} approaches, 
accurate numerical results on the phase diagram of the quantum model 
have become available only recently \cite{mc3d,kok}.  
Since the pioneering works based on spin-wave theory \cite{oguchi} 
and on high-temperature expansions \cite{domb}, a better understanding of 
the behavior in two dimensions \cite{mc2d} and of the critical 
properties in three dimensions \cite{mc3d,chen} has been provided 
by more and more efficient MC methods.
Nevertheless, several  problems are still open, especially in the 
low and intermediate temperature region. 
This regime has been successfully investigated by means of the RG,  
after an appropriate mapping onto the nonlinear  sigma model 
(\nlsm) \cite{chn}. 
The satisfactory agreement between RG and experimental 
results, however,  seems to hold only for spin $1/2$ and   
quantitative deviations between theory and 
experiments \cite{birgeneau1} appear for larger values of $S$.
Both Monte Carlo simulations \cite{mc2d} and series expansions \cite{elstner}
suggest that the breakdown of the \nlsm description of a Quantum
Antiferromagnet is due to the intrinsic limitations of the methods 
based on effective actions. In fact, these approaches 
are devised for describing the long wavelength behavior of the model and
can be justified in a regime characterized by strong correlations, 
while the experimental investigations span a wide temperature range, including 
regions where local antiferromagnetic order is weak.
The  same conclusions are also found by use of a new semiclassical approach
in which the problem of the quantum \HH antiferromagnet is reduced to a classical
one at an effective temperature taking into account quantum fluctuations
\cite{tognetti}.

Besides the problems concerning the \HH antiferromagnets in bipartite lattices,
several other questions are still open in the framework of phase 
transitions in quantum systems. Generally, accurate 
analytical approaches at finite temperature are not available and
quantum simulations are often hindered by the ``sign problem".
For instance, frustrated antiferromagnets or fermionic systems, like  
the celebrated Hubbard model, are still beyond the 
capabilities of today's simulation algorithms. Interestingly, this class of 
models might be investigated by a straightforward 
generalization of the method we are going to present here,
at least in order to estimate the magnetic phase boundaries or the 
occurrence of phase separation.

In this paper, we present a new approach to the phase transitions 
in quantum systems. A preliminary account can be found in Ref. \cite{pla}.
We develop a general formalism which
can be used in a wide range of temperatures both above and 
below a possible critical point. The method is then applied to the specific 
case of \HH antiferromagnets, for different values of the spin and of 
the spatial dimensions. This approach, which we called
QHRT (Quantum Hierarchical Reference Theory) is a quantum extension of 
the classical HRT which was developed in the context of liquid-vapor transitions 
and successively applied to several system, like Ising models
and binary mixtures \cite{hrt}. Our method relies on the fact that, for a large 
class of quantum systems, an exact mapping exists between the quantum 
partition function and a fictitious classical one to which HRT can be applied.
The main advantage of HRT, also shared by QHRT, is its microscopic character,
which allows to obtain both universal and non universal quantities of 
the system under consideration. Non classical critical exponents 
together with a precise determination of the critical temperature 
and a correct description of the first order phase transitions
are in fact obtained within the QHRT formalism.
As already mentioned, the theory is very general and 
can in principle be applied to
several physical systems, bosons, fermions and spins.

Following the idea of the RG approach, an evolution equation
is derived, starting from the perturbative expansion of the Helmholtz free energy.
Fluctuations over progressively larger lengthscales are taken into account
by varying a cut-off parameter thereby connecting the system with
no fluctuations (mean-field approximation) to the fully interacting one.
The equation obtained in this way is formally exact, although it is not written 
in a closed form. In order to get a closed equation, the knowledge of the 
structure of the dynamic 
correlation functions of the model as a function of frequency and wavevector
is needed. This represents the only approximation present in the QHRT 
approach and requires a thorough analysis of its physical implications.
Near a critical point our partial differential equation acquires a 
universal form and reproduces the RG structure for a 
three component order parameter with critical exponents exact to 
first order in $\e=4-d$. Also as $T\to 0$, the QHRT approach is able to 
predict the known behavior originally derived by a RG analysis of the 
\nlsm.  
We solved the equation for several values of the spin, both in two 
and three spatial dimensions. In $d=3$ we obtained an excellent matching with
the available numerical simulation data, also from a quantitative point of view.

The paper is organized as follows: In Section \ref{2} we give the derivation of
the theory. In Section \ref{3} the theory is applied to the \HH 
antiferromagnets, the
classical \HH model being obtained as the limiting $ S\ra \infty$ case.
In Section \ref{4} an explicit closure is given and the implications of 
this choice are
discussed. In Section \ref{5} an analysis of QHRT equation in proximity 
of a phase
transition is given. In Section \ref{6} the explicit results, obtained 
via numerical
integration  of the evolution equation, are presented. Finally 
in Section  \ref{conclusions} conclusions are drawn and
further possible applications of our theory are proposed.

\section{Derivation of QHRT equations} \label{2}

\nn 
In this Section we develop a general formalism for the description of
critical phenomena in quantum systems. Our goal is to implement the
basic ideas of renormalization group to the microscopic, many body
hamiltonian of an interacting quantum model. 
The class of hamiltonians we are going to consider, in 
arbitrary dimension $d$, has the structure:
\be 
\H=\H_R+\V=\H_R +{1\over 2}\int dx dy {\rrr(x)} w(x-y) {\rrr(y)} \label{hhh} 
\ee
\nn 
where $\H_R$ is a reference part whose properties are
supposed known and the coordinates $x$ and $y$ span the 
$d$-dimensional space. $\V$ represents an interaction term
which couples the (local) observables $\rrr(r)$ at two
different points via the translationally invariant 
two body potential $w(x-y)$. In the following we will
assume that: $i)$ the local operators $\rrr(r)$ commute
at different points and $ii)$ the interaction $w(r)$
can be Fourier transformed. Instead, no limitation
is imposed on the form of $\H_R$, which is {\it not} restricted to 
describe free particles. 
This class of hamiltonians is indeed rather general and
includes several systems of widespread interest, like
quantum magnets (where $\rrr(r)$ represents a spin variable)
but also models of interacting fermions, both on and off
lattice (where $\rrr(r)$ is the local density operator) 
and even systems with electron-phonon coupling, like the Hubbard-Holstein model.
Generalizations to other cases, for instance models with 
many particle interactions, are possible but will not be considered here.

As a first step, we will write the partition function of the quantum
model in a form identical to that of a classical system.
The latter will be then studied by means of the powerful methods already 
available in the framework of classical critical phenomena. 
This {\sl quantum to classical} mapping is most easily carried out
by expanding the grand canonical partition function $\Xi$
as a power series of $w(r)$. The method we use closely follows
the standard treatment of the path integral formulation of 
quantum mechanics \cite{feynman}, although we carefully avoid
to introduce any smoothness assumption on the classical fields.
Therefore, our treatment is {\it not} limited to
long wavelength or low energy and  the equations we are
going to obtain are formally exact in the whole phase diagram
of the model.

The grand partition function of the quantum system,
$ \Xi ={\rm Tr}\,\exp(-\b \H)$, is first written in terms of the
(imaginary time) evolution operator \cite{fetter} as:
\be 
{\rm Tr}\,\exp(-\b \H) = {\rm Tr}\,\{\exp(-\b \H_R) \hat U(\b) \} 
\label{part00}
\ee 
\nn 
where $\hat U(t)$ is the solution of the differential equation:
\be
\frac{d}{d t}  \U (t) = -  \V(t) \U (t) 
\label{e1} 
\ee
\nn 
with $\U(0) = 1$ as initial condition.
$\V(t) $ is the time dependent operator:
\be 
\V(t) =  \exp[\H_R t] \, \V \, \exp[-\H_R t] 
\ee
\nn  
Equation (\ref{e1}) can be formally solved by iteration with the result:
\be 
\hat U(t)=\sum _{n=0}^{\infty} \frac{(-1)^n}{n!} 
\int _{0}^{t}dt_1 ... \int _{0}^{t}dt_n T_{t} [\V(t_1) ... \V(t_n) ] 
\label{soluzione} 
\ee
\nn 
where $T_t$ is the usual time ordering operator.
By inserting in (\ref{soluzione}) the explicit form of $\V(t)$
\be
\V(t) = \frac{1}{2} \int dx dy \int dt' 
\rrr (x,t) w(x-y)\d(t-t') \rrr (y,t')
\ee
\nn 
equation (\ref{part00}) can be written as a power series of
the $(d+1)$-dimensional interaction
\be    
\f (X,X')=w(x-x')\delta(t-t') 
\label{w} 
\ee 
Here, the variable $X$ identifies the pair $(x,t)$ where $x$ spans
the $d$-dimensional coordinate space and $t$ belongs to the
interval $(0,\b)$.
The explicit expression for the grand partition function $\Xi$ is:
\beqa  
 \Xi &=& {\rm Tr} \,e^{-\b \H} \nm \\
&=& \Xi_R \,\sum_{n=0}^{\infty} \frac{(-1)^n}{2^n\, n!}\, \int \, dX_1 dX_1'
...
dX_n dX_n' \, \r _R(X_1,...X_n')\,\f (X_1,X_1') ...\,\f (X_n,X_n') 
\label{z} 
\eeqa
\nn 
Here and in the following $R$ labels the reference system defined by $\H_R$.
The imaginary time dynamic correlation functions appearing in Eq. 
(\ref{z}) $\r(X_1,...X_n)$ are defined, for a general hamiltonian $\H$, by:
\be  
\r (x_1,t_1;...x_n,t_n)= \langle \rrr (x_1,t_1) ...\rrr (x_n,t_n) \rangle
={1\over\Xi}
{\rm Tr}\; \left\{ e^{-\beta \H} \,T_t [\rrr (x_1,t_1) ...\rrr (x_n,t_n)] 
\right\}
\label{correlazioni} 
\ee
\nn 
where $\rrr(x,t)$ is the local density operator in imaginary time: 
\be 
\rrr(x,t) =  \exp[\H t] \, \rrr(x) \, \exp[-\H t] 
\label{timedep}
\ee
\nn  
The correlation functions in the perturbative expansion (\ref{z}) 
are given by Eqs. (\ref{correlazioni},\ref{timedep}) specialized to the
reference hamiltonian $\H_R$.
Since $\rrr(r)$ commutes on different sites, the functions 
(\ref{correlazioni}) are symmetric under the permutation of the 
labels and can be viewed as the correlation functions 
of a hypothetical $(d+1)$ dimensional {\sl classical} reference system.
The quantum nature of the problem is then contained in the structure of
the reference system through $\Xi_R$  and $\r_R(X_1,...X_n)$. 
It is interesting to notice that the 
n-particle correlation functions $\r (X_1,...X_n)$
associated to a given quantum hamiltonian $\H$ can be equivalently obtained
as successive functional derivatives of a generalized partition function
$\Xi[J]$ with respect to a space and time dependent external field
$J(x,t)$ coupled to the quantum operator $\rrr(x)$. 
More precisely, the generating functional $\Xi[J]$ is still
defined as the trace of an imaginary time evolution operator:
\be
\Xi[J]={\rm Tr} \ll \{ \exp[-\b \H]\U(\b)\rr \}
\label{genfun}
\ee
where now $\U(t)$ is the solution of the differential equation:
\be
\frac{d}{d t}  \U (t) = - \hat K (t)  \U (t)   
\label{e2} 
\ee
\nn 
with initial condition $\U(0) = 1$ and
\be
   \hat K (t)=  -\int dx    J(x,t) \rrr(x,t)
\ee 
\nn 
Following the same steps which led to Eq. (\ref{z}) it is easy to show
that the correlation functions $\r(x_1,t_1;...x_n,t_n)$ generated 
by $\Xi[J]$ 
\be  
\r (x_1,t_1;...x_n,t_n)=\ll  . \frac{1}{\Xi[J] }\;\frac{\d ^n\Xi[J] }
{\d J(x_1,t_1)...\d J (x_{n},t_{n})}
\rr  | _{J=0 }
\label{rn}
\ee  
reduce to those defined in (\ref{correlazioni}). Analogously, it is useful 
to introduce the connected correlation functions defined by:
\be  
F (x_1,t_1;...x_n,t_n)=\ll  . \frac{\d ^n\ln\Xi[J] }
{\d J(x_1,t_1)...\d J (x_{n},t_{n})}
\rr  | _{J=0 }
\label{fn}
\ee  
In the following, we will take the
perturbative expansion (\ref{z})  as a definition of the fully interacting 
partition function where the reference n-particle correlation functions are
written as functional derivatives via  Eqs. (\ref{rn},\ref{fn}). 
\nn
Now the quantum to classical correspondence is complete: 
Having defined the (quantum) reference system by a hamiltonian $\H_R$,
we compute the corresponding functional $\Xi_R[J]$ via Eq. (\ref{genfun}).
If we interpret this quantity as the grand partition function of a fictitious
classical reference system in $d+1$ dimensions in 
an external field $J(x,t)$ coupled to a scalar order parameter $\psi(x,t)$
and at an effective temperature $T_{cl}=1$,
the exact partition function of the quantum system (\ref{z}) 
formally coincides with that of such a classical system 
with the additional non local coupling (\ref{w}) in the field $\psi(x,t)$
at two different space-time points.

For future reference, let us briefly discuss the general relationship 
between the previously introduced imaginary time two point function
(\ref{rn}) and the physically relevant dynamic correlations defined by:
\be  
D(x_1-x_2,t_1-t_2)= {i\over\Xi}\,
{\rm Tr}\; \left\{ e^{-\beta \H} \,e^{i \H t_1}\Delta\rrr (x_1) e^{-i \H t_1}
e^{i \H t_2}\Delta\rrr (x_2)e^{-i \H t_2} \right\}
\label{realt} 
\ee
\nn 
where $\Delta\rrr(x)=\rrr(x)-<\rrr(x)>$.
Taking the Fourier transform in the space and time variables 
and inserting a complete set of basis states we get the usual Lehmann
representation for the spectral function:
\be
{\rm Im}\,D(k,\om)={2\pi\over\Xi}\, \sum_{s,s^\prime} \,e^{-\beta E_s} 
\,\vert <\,s\,|\rrr_k|\,s^\prime>\vert^2 \,\delta(\om-E_{s^\prime}+E_s)
\label{spec}
\ee
\nn
which obeys the detailed balance relation ${\rm Im}\,D(k,\om)=
{\rm Im}\,D(-k,-\om) \exp(\b\om)$. 
The analogous expression for the imaginary time correlation function
reads:
\be
F(k,\om_n)={1\over\Xi}\, \sum_{s,s^\prime} \,{e^{-\beta E_s} 
-e^{-\beta E_{s^\prime}}\over E_{s^\prime}-E_s-i\om_n}\,
\vert <\,s\,|\rrr_k|\,s^\prime>\vert^2 
\label{lehm}
\ee
\nn
where we took advantage of the periodicity of the two
particle correlation function (\ref{correlazioni}) in
the imaginary time variable $t=t_1-t_2$ with period $\b$.
As a consequence, the frequency variable $\om_n$ is now an integer
multiple of $2\pi/\b$.
By comparing Eqs. (\ref{spec}) and (\ref{lehm}) we get the 
final relation, valid for systems with inversion symmetry:
\be
F(k,\om_n)=\int_{-\infty}^{\infty} {d\om\over\pi}\, {\rm Im}\,D(k,\om)
{\om\over \om^2+\om_n^2} \qquad\qquad {\rm for} \;\;\om_n\ne 0
\label{rela}
\ee
\nn
This equation will be useful in order to relate the correlation
function $F(k,\om)$ to physical observables.

In order to derive a set of renormalization group equations for
the model hamiltonian (\ref{hhh}), we follow the same strategy 
successfully adopted in
classical systems \cite{hrt}: First we obtain a formal expansion of 
the Helmholtz free energy of the quantum model in powers of the
two body interaction $w(x)$. Then we define a sequence of 
intermediate systems ({\it Q-systems}) in which fluctuations
characterized by wavevectors $k<Q$ are inhibited. Finally we
write a differential equation for the ``evolution" of the 
free energy when the infrared cut-off $Q$ is changed by an infinitesimal
amount. Analogously to the classical case, the resulting equation 
is exact but cannot be written in closed form because it involves
the knowledge of the two particle correlation function of the $Q$-systems. 
In fact, this is only the first equation of an infinite hierarchy of
differential equations for the free energy and n-particle correlation
function of the model, which form the Quantum Hierarchical Reference 
Theory of Fluids (QHRT). 

Let us briefly sketch the basic steps necessary to derive the first 
QHRT equation: Details can be found in Ref. \cite{hrt} via
the quantum to classical correspondence previously discussed.
The perturbative expansion of the logarithm of the
grand partition function is directly obtained from Eq. (\ref{z}) 
in terms of the connected correlation functions (\ref{fn})
of the reference system, denoted by $F_R$.
Next we perform a Legendre transformation which defines the Helmholtz 
free energy $A[\rho]$ as a functional of the order parameter 
profile $\r(x)=\b^{-1} \left ( \d \ln \Xi / \d J(x) \right )$ 
corresponding to a static external field $J(x)$:
\be 
A[\r] = -\b^{-1} \ln \Xi[J] + \int dx  J(x) \r (x)  
\label{legendre}
\ee
The perturbation series of the free energy in powers of the 
interaction is then given by a diagrammatic expansion 
involving the pair potential (\ref{w}) and the
reference correlation functions (\ref{fn}). This series can
be conveniently ordered according to the number of loops: 
the terms up to the one loop level reproduce the well known 
random phase approximation (RPA) to the free energy of the 
interacting system. In the homogeneous limit $\rho(x)=\rho$,
the expansion reads:
\be 
A = A_R + \frac{1}{2} V \rho ^2 \ww(k=0)- \frac{1}{2} V \rho w(r=0)
+ V {1\over 2\b} \sum_{\om_n} \int \frac {d^dk}{(2 \pi )^d }
\ln \ll [ 1+F_R(k,\om_n) \ww(k) \rr ] + { \cal L}_2
\label{sviluppo} 
\ee
Here $V$ is the volume of the system,
${\cal L}_2$ represents the sum of all the diagrams with at least two loops,
whose structure we do not need to detail, $k$ is the wavevector
and $\om_n=2\pi n/\b$ are the usual Matsubara frequencies
coming from the Fourier transform in the imaginary time direction,
while $\ww(k)$ is the Fourier transform of the two body 
interaction $w(x)$.

Now we define a sequence of $Q$-systems interpolating between 
the reference system and the fully interacting one by introducing
a cut-off $Q$ in the wavevector integration of every loop of the expansion 
(\ref{sviluppo}). Clearly, for $Q\to 0$ we
recover the full expansion (\ref{sviluppo}) to all orders
while for $Q\to\infty$ all contributions to the free
energy with at least one loop are suppressed and we get the 
mean field expression for the Helmholtz free energy:
\be
A_{Q=\infty} = A_R +  \frac{1}{2} V \rho ^2 \ww(k=0)- \frac{1}{2} V \rho w(r=0)
\label{mfa}
\ee
Analogously, the two point function $F(k,\om_n)$ admits a perturbative
expansion in powers of $w(r)$ which can be again ordered according to the
number of loops. Only the zero loop (chain) terms survive 
in the $Q\to \infty$ limit reproducing the well known RPA result:
\be
F_{Q=\infty}(k,\om_n) = {F_R(k,\om_n)\over 1+F_R(k,\om_n) \ww(k)}
\label{mff}
\ee
\nn
These results show that, through the sequence of $Q$-systems, 
thermal and quantum fluctuations are gradually turned on and
consequently, the free energy of the $Q$-system continuously
connects the mean field approximation (\ref{mfa},\ref{mff}) 
to the exact expression 
corresponding to the fully interacting model.

As already noticed for the classical case \cite{hrt},
it is possible to write an exact differential
equation describing the $Q$-evolution of the free energy $A_Q$.
The key observation is that in all the diagrams of the expansion
(\ref{sviluppo}), including the terms collectively named ${\cal L}_2$,
every loop contains at least an interaction bond. Therefore, 
limiting each momentum integration to the domain $k>Q$ can be equally
achieved by cutting off the Fourier components of the interaction
$\ww(k)$ at small momenta. Then, the $Q$-system can be equivalently
defined as the system interacting via the pair potential:
\be
\ww_Q(k) = \cases{
{\displaystyle
\ww(k)} & if $k > Q$ \cr
{\displaystyle
0} & if $k < Q$ \cr}
\label{deltaw}
\ee
\nn
The only {\sl caveat} concerns the zero loop diagrams in the diagrammatic
expansion of the free energy, which only depend on the zero momentum
component of the perturbation potential $\ww_Q(k)$. According to the
definition (\ref{deltaw}), $\ww_Q(k=0)$ identically vanishes for every 
$Q\ne 0$: The zero loop contributions are then absent from the 
free energy corresponding to the potential (\ref{deltaw}), while were 
included in the previous definition of $Q$-system. Therefore, in order
to get the free energy of the $Q$-system from the free energy of the
hypothetical model interacting through the potential $w_Q$, we must 
explicitly add the zero loop contributions, which have been already 
evaluated in Eq. (\ref{mfa}). Similar considerations hold for the
two point function of the $Q$-system $F_Q(k)$ which differs from that 
corresponding to the interaction $w_Q$ because of the simply connected
(chain) diagrams. However, these terms
are proportional to the $k^{th}$ Fourier component of the interaction $\ww_Q(k)$
which, according to the definition (\ref{deltaw}), coincides with $\ww(k)$
for every $k>Q$ while vanishes if $k<Q$. Therefore, the chain
contribution needs to be added to the two particle correlation function
of the system with interaction $w_Q$ only for $k<Q$. As a consequence, 
the two point function of the model with interaction $w_Q$, 
which we denote by $\hat{F}_Q(k,\om_n)$ is discontinuous in $k$ space 
on the $k=Q$ surface and can be expressed in terms of the 
two point function of the $Q$-system $F_Q(k)$ as 
\be
\hat{F}_Q(k,\om_n) = \cases{
{\displaystyle
F_Q(k,\om_n)} & if $k > Q$ \cr
{\displaystyle
{F_Q(k,\om_n)\over 1-F_Q(k,\om_n)\ww(k)}} & if $k < Q$ \cr}
\label{fbare}
\ee
\nn
This one to one correspondence between $Q$-systems and models
interacting via the potential (\ref{deltaw}) is extremely useful in 
order to obtain the QHRT evolution equations. According to the previous
discussion,
the change in the free energy of the $Q$-system corresponding to a change 
in the cut-off $Q\to Q-\delta Q$ can be evaluated by means of the perturbative
expansion (\ref{sviluppo}) in which the ``reference" free energy is now
identified as $A_Q$ and the perturbation potential as 
$ \left [\ww_{Q-\delta Q}(r)- \ww_Q(r)\right ]$.
Such a potential has Fourier components only in the infinitesimal
momentum shell defined by $Q-\delta Q <k< Q$. In the limit $\delta Q\to 0$,
by keeping only terms $O(\delta Q)$ we just select the one loop diagrams
which have been explicitly summed up in Eq. (\ref{sviluppo}), while 
all the many loop contributions ${\cal L}_2$ are at least of order 
$(\delta Q)^2$, i.e. rigorously negligible. The zero loop terms 
are trivial and have been automatically included form the 
beginning through the initial condition (\ref{mfa}).
The resulting differential equation for the evolution of the
free energy density is then:
\be 
\frac{d}{dQ} \ll ( \frac { {{A}}_Q }{V} \rr )=
{Q^{d-1}\over 2\beta} \sum_{\om_n} \int _{k=Q} \frac {d \Omega _k}{(2 \pi)^d}
\ln [1- F_Q(k,\om_n) \ww(k) ] 
\label{chrt}
\ee
\nn
where the momentum integral is limited on the shell $k=Q$ and then
reduces to the integration over the solid angle $\Omega _k$.

This evolution equation is exact and must be supplemented by 
the initial condition (\ref{mfa}) at $Q=\infty$. 
However, the presence of the two point function
$F_Q(k,\om_n)$ at right hand side prevents the possibility to 
obtain the thermodynamics of the model without introducing
some sort of approximation. In the following we will discuss
a closure to Eq. (\ref{chrt}) based on a parametrized form of
the two point function which is related by an exact sum rule
to the Helmholtz free energy $A_Q$ of the $Q$-system.
The justification of this closure relation requires, however,
an analysis of the physical properties of the model we want
to investigate and therefore we first have to specialize this
rather general presentation of the QHRT equations to the case
of the antiferromagnetic Heisenberg model which will
be thoroughly investigated in the following Sections.

\section{QHRT equation for the isotropic Heisenberg  model} \label{3}

\nn In this Section, the QHRT formalism just developed 
will be applied to the study of the isotropic antiferromagnetic
\HH model on a $N$-site cubic lattice in arbitrary dimension $d$ and
spin $S$. This is an extremely interesting many body model which 
allows to study the interplay of thermal and quantum fluctuations 
in determining the
phase diagram of the system and the effects of spatial dimensionality on the 
occurrence of long range order \cite{mermin}. In the 
$S\to \infty$ limit, quantum fluctuations are suppressed and the
system behaves classically. This regime has been extensively studied by 
series expansions \cite{domb}, numerical simulations \cite{binder},
renormalization group techniques \cite{zinn} and approximate theoretical
approaches \cite{manousakis}. Much less is known at finite values of the
local spin $S$: accurate simulations have become available 
only recently while series expansions are known 
to lose accuracy both at low temperature and in the critical region. 

Since we want to investigate the paramagnetic-antiferromagnetic transition,
it is convenient to consider the presence of an external magnetic 
field $h$, parallel to the $z$ axis, which couples to the order parameter, 
i.e. to the staggered magnetization:

\be
\H=
J \sum_{<\bR,\bR^\prime>} {\bf \S_R}\cdot {\bf \S}_{{\bf R}^\prime}
-h \sum_\bR e^{i\bg\cdot\bR} \S^z_\bR
\label{hheis}
\ee

\nn Here $J>0$,  ${\bf g}$ is
the antiferromagnetic wavevector of components $g_i=\pi$, while 
the spin operators ${\bf \S}$ satisfy the usual commutation relations

\be 
\ll [\S_{{\bf R}}^{i},\S_{{\bf R}^\prime}^{j} \rr ] = i \d_{{\bf
R,R}^\prime} \epsilon _{i \, j \, k} \S_{\bf R}^{k} 
\ee

\nn The lower and upper indices label respectively
the site on the lattice and the component of the spin, while
$\epsilon _{i \, j \,k } $ is the totally antisymmetric tensor.
With respect to the class of hamiltonians examined in the previous
Section, the \HH model requires some straightforward extension
because $i)$ the system is defined on a lattice, which is not isotropic;
$ii)$ the order parameter is now the staggered magnetization:
a three component vector. As a consequence, the correlation functions
acquire spin component indices and $iii)$ the presence of a staggered 
magnetic field breaks translation invariance by one lattice site.

The reference system is chosen to be defined by the non interacting
hamiltonian

\be
\H_R=-h \sum_\bR e^{i\bg\cdot\bR} \S^z_\bR
\ee
\nn
whose properties can be easily obtained. For instance, the 
grand potential as a function of the number of sites $N$, inverse
temperature $\b$ and staggered field $h$ reads:

\be
\log\Xi_R=N\ln \left [ \sinh(\b h S)\coth(\b h /2) + \cosh(\b h S)\right ]
\label{init}
\ee
\nn
The staggered magnetization $m$ can be found by differentiating (\ref{init})
with respect to $\b h$ and the Helmholtz free energy then follows by
Legendre transform (\ref{legendre}). However, the explicit inversion 
formulae can be found analytically  only in the $S=1/2$ case where

\beqa
\b A_R &=& N
\left [ (\frac{1}{2} +m )\ln
(\frac{1}{2} +m )+
(\frac{1}{2} -m )\ln (\frac{1}{2} -m ) \right ] \nm\\
\b h &=& \ln\left [ {\frac{1}{2} + m\over \frac{1}{2} - m}\right ]
 \label{reference}      
\eeqa
\nn
In order to derive the renormalization group equation 
appropriate for the magnetic transition 
in this model, we now have to suitably define the sequence of $Q$-systems 
whose role is to gradually introduce fluctuations into the system.
The critical fluctuations occur at a wavevector $\bk=\bg$ and 
then it is natural to define a sequence of systems in which the 
Fourier components of the potential $\ww(\bk)$
\be
\ww(\bk)= 2 J \g (\bk)  \qquad \qquad  \g  (\bk)=\sum_{i=1} ^d \cos k_{i}
\ee  
\nn
are suppressed in a neighborhood of $\bk=\bg$. A convenient choice is
to limit the domain of integration in $\bk$ space by
\be
|\g(\bk)|<\sqrt{d^2-Q^2}
\label{defq}
\ee
\nn
with $Q\in (0,d)$. As $Q\to 0$ all Fourier components are included
and we recover the fully interacting system.
At $Q=d$ the free energy density $a=A/N$ includes only the reference term and 
the zero loop contribution:
\be
a_{Q=d}= a_R -dJ m ^2
\label{inizio}      
\ee
The two particle correlation function is represented by a
a $3 \times 3$ matrix which, due to the spin 
symmetries in the Hamiltonian under rotations around the $z$ axis, 
has the following general structure in Fourier space: 

\beqa
\nonumber \\
\nonumber \\
F=\ll ( 
\begin{array}{lll}
 F^{xx} &  F^{xy} & 0 \\ 
F^{yx} &  F^{yy} & 0 \\ 
0 & 0 &  F^{zz} \label{F2}
\end{array}
\rr ) \\
\nonumber \\
\nonumber
\eeqa
\nn The three independent entries are given by:
\beqa
{F}^{xx}(\bk _1,\om _1;\bk_2,\om_2) 
&=& 
{F}^{yy}(\bk _1,\om _1;\bk_2,\om_2) =
F^{xx}(\bk_1,\om_1)\, {\b}\,N \d ({\om} _1+ {\om} _2)\,
\d(\bk_1+\bk_2) \nm  \\
\nm \\
{F}^{xy}(\bk_1,\om   _1;\bk_2,\om _2)
&=&
-{F}^{yx}(\bk_1,\om   _1;\bk_2,\om _2)=
F^{xy}(\bk_1,\om_1)  \, {\b }\, N \d ({\om } _1+ {\om } _2) 
\d (\bk_1+\bk_2- \Q )
\nm   \\
\nm \\
{F}^{zz}(\bk_1,\om _1;\bk_2,\om _2)&=&
F^{zz}(\bk_1,\om_1)  \, {\b }\, N \d ({\om } _1+ {\om} _2)\, 
\d (\bk_1+\bk_2) \label{Fij}  
\eeqa
\nn 
where $\bk$ belongs to the first Brillouin zone and  
$\om$ is an integer multiple of $2\pi /\beta$.  
Because of the presence of the external field, the translational invariance 
is broken and the two point function is not proportional to 
$\d(\bk_1+\bk_2)$. 
Nevertheless, due to the symmetry of the hamiltonian under a 
translation of one 
lattice site and a simultaneous rotation of $\pi$ around $x$ axis,  
the diagonal elements and the off-diagonal elements are respectively 
proportional to $\d(\bk_1+\bk_2)$ and to $\d(\bk_1+\bk_2- \Q )$
as stated in Eq. (\ref{Fij}).
At $Q=d$, the summation of the zero loop diagrams explicitly gives:
  
\beqa
&& F^{xx}_{Q=d} (\bk,\omega) ={\mu_\perp-\ww(\bk) \over  m^{-2}\omega^2+
\mu_\perp^2-\ww(\bk)^2}   \nm  \\
&&  \nm \\
&& F^{xy}_{Q=d}(\bk,\omega)={m^{-1}\omega\over
m^{-2}\omega^2+\mu_\perp^2-\ww(\bk)^2}  \nm \\ 
&& \nm \\
&& F^{zz}_{Q=d}(\bk,\omega)={\delta_{\omega,0}\over\mu_{||}+\ww(\bk)}  
\label{fij}
\eeqa 
\nn
where $\mu_\perp$ and $\mu_{||}$ are functions of temperature and 
staggered magnetization $m$ given, for $S=1/2$, by:
 
\be
\b\mu_\perp=2m^{-1}\tanh^{-1}(2m) \qquad\qquad
\b\mu_{||}=4(1-4m^2)^{-1}
\label{mu0}
\ee
\nn
Having determined the mean field form (i.e the zero loop contribution)
of the relevant quantities, we now turn to the inclusion of fluctuations
via the definition of the QHRT renormalization group equation.
Following the procedure outlined in the previous Section 
and taking into account the definition (\ref{defq}) of $Q$-systems,
we find the exact evolution equation for the free energy density
$a_Q=A_Q/N$ of the antifferromagnetic \HH model which should 
be solved with the appropriate initial condition (\ref{inizio}) at $Q=d$: 

\beqa  
&&\frac{d \, a_Q}{dQ}=\frac{1 }{2\b} 
\sum_{\om_n}
\int {d^d k\over (2\pi)^d}\,\, \d(\sqrt{d^2-\g(\bk)^2}-Q)
\,\Big \{ \ln (1-F^{zz} _Q(\bk,\om_n)\ww(\bk)) +
\label{qhrt} \\
&& \ln \left [ (1-F^{xx} _Q(\bk,\om_n)\ww(\bk))
(1+F^{xx} _Q(\bk+\bg, \om_n)\ww(\bk))- 
F^{xy} _Q(\bk,\om_n)F^{xy} _Q(\bk+\bg,\om_n)  \ww(\bk)^2 \rr] 
\Big \}\nm
\end{eqnarray}

\nn The summation of the one loop (ring) diagrams, necessary 
to obtain Eq (\ref{qhrt}), must be carried out
by carefully exploiting the frequency and momentum conservation
rules (\ref{Fij}). The delta function present in the
$d$-dimensional integral (\ref{qhrt}) limits the 
integration  domain to the
surface $\Sigma_Q$ defined by $\gamma(\bk)=\pm\sqrt{d^2-Q^2}$. 
This equation, although
formally exact, is not closed because the evolution of the free energy
depends on the unknown magnetic structure factors of the model 
$F^{ij}_Q(\bk,\omega)$ at a generic value of the cut-off $Q$
while these quantities are known only at the beginning of the
integration, i.e. at $Q=d$ where they have the form shown in Eq. (\ref{fij}).
If we simply assume that $F^{ij}_Q(\bk,\omega)$ is not affected by 
fluctuations, the QHRT equation reproduces the the standard RPA
approximation. A better treatment of fluctuations needs a more
flexible parameterization of the correlation functions.
This delicate point requires a detailed analysis of the 
properties of  the (imaginary time) dynamic structure factors
of the \HH model, which will be carried out in the next Section.
\section{Two body correlations} \label{4}

\nn 
In this Section we exploit some known feature of quantum antiferromagnets 
in order to derive constraints on the formal structure of 
the two point functions $F^{ij}_Q(\bk,\om_n)$
required to close the QHRT evolution equation (\ref{qhrt}).

\nn
As a first step, let us recall the relationship between the 
dynamic correlations of the \HH model, defined in physical
frequency, and the imaginary time functions which enter the QHRT
formalism. The general correspondence has been already obtained
in Eq. (\ref{rela}): Here we point out that the analytical form
of the mean field correlation functions (\ref{fij}), which serves as 
initial condition, reproduces the 
single mode approximation in the dynamic structure factor:
\beqa
{\rm Im} \,S^{xx}(\bk,\om) &=& {\pi\om\over 1-\exp(-\b\om)}
{\delta(\om-\epsilon_k) + \delta(\om+\epsilon_k)\over \mu_{\perp}+
\ww(\bk)}\nm\\
{\rm Im} \,S^{zz}(\bk,\om) &=& {2\pi\over\b}\, 
{\delta(\om) \over \mu_{||}+\ww(\bk)}
\label{singlem}
\eeqa
where 
\be
\epsilon_k = m \sqrt{\mu_{\perp}^2-\ww(\bk)^2}
\label{disp}
\ee
\nn
can be recognized as the dispersion relation for spin waves in this
approximation.
The elastic sum rule immediately gives the corresponding form of 
the equal time structure factor via integration over the frequency axis:
\beqa
S_{\perp}(\bk)&=& {1\over \b}\sum_{\om_n} F^{xx}(\bk,\om_n) =
{\epsilon_k\over 2\tanh(\b\epsilon_k/2)}\,{1\over \mu_{\perp}+\ww(\bk)}\nm\\
S_{||}(\bk)&=& {1\over \b}\sum_{\om_n} F^{zz}(\bk,\om_n) =
{1\over\b}{1\over \mu_{||}+\ww(\bk)}
\label{static}
\eeqa
These expressions show that both the dynamic and static structure 
factors (\ref{singlem},\ref{static}) coincide with those usually 
adopted in the interpretation of neutron scattering data, 
in the limit of negligible spin wave damping \cite{tyc}.
This suggests that the simple analytical structure of Eqs. (\ref{fij})
qualifies as a good starting point for obtaining a closure
relation to the QHRT equation. However, the parameters 
$\mu_{\perp}$ and $\mu_{||}$ appearing in the correlation functions
(\ref{fij}) will be modified by fluctuations and cannot be identified
with their mean field estimate (\ref{mu0}) down to $Q=0$. These quantities
can be directly related to important physical properties 
of the model: The definition (\ref{fn}) of n-point function 
shows that the zero frequency value of $F^{ii}(\bk,\om_n)$ coincides with 
the uniform (staggered) susceptibility $\chi^{\circ}$
($\chi^s$) when $\bk=0$ ($\bk=\bg$) and therefore
\beqa
\chi^{\circ}_{\perp}&=&F^{xx}(0,0)=(\mu_\perp + 2Jd)^{-1} \nm\\
\chi^{\circ}_{||}&=&F^{zz}(0,0)=(\mu_{||} + 2Jd)^{-1} 
\label{homo} \\
\chi^s_{\perp}&=&\left({h\over m}\right)^{-1}=
m\left ({\partial a\over\partial m}\right )^{-1} =
F^{xx}(\bg,0)=(\mu_\perp - 2Jd)^{-1} \nm\\
\chi^s_{||}&=&\left ({\partial h\over \partial m}\right)^{-1}=
\left ( {\partial^2 a\over \partial m^2 }\right)^{-1}=
F^{zz}(\bg,0)=(\mu_{||} - 2Jd)^{-1} 
\label{stag}
\eeqa
Here, we have conveniently expressed the staggered susceptibilities 
as thermodynamic derivatives of the external field $h$ and of the free energy
density $a$ with respect to the staggered magnetization
$m$, using the isotropy of the model in the spin variables.
At the critical point and in the whole coexistence region, both
the longitudinal and the transverse susceptibilities diverge \cite{zinn},
implying that $\mu_\perp=\mu_{||}=2Jd$. This feature leads to a gapless
spin wave spectrum, via Eq. (\ref{disp}), with linear behavior 
near $\bk=0$ (and $\bk=\bg$): $\epsilon_k\sim c_s k$. 
In the approximation (\ref{fij}), the associated spin velocity is
proportional to the spontaneous staggered magnetization: 
$c_s=Jm\sqrt{4d}$. At zero temperature, the
static transverse structure factor (\ref{static}) is linear in $k$
around $\bk=0$ in the symmetry broken phase and the prefactor defines 
the spin stiffness $\rho_s$ \cite{stringari}:
\be
S_\perp(k)\sim {\rho_s \over 2c_s} k
\label{stiff}
\ee
\nn
Within a parameterization of the form
(\ref{fij}), the stiffness can be expressed in terms of the spin velocity: 
$\rho_s= c_s^2/(4Jd)=Jm^2$. 
Remarkably, the hydrodynamic relation $\chi^{\circ}_\perp c_s^2=\rho_s$, 
which is believed to hold at zero temperature, is satisfied even
in this simple approximation. 

A known sum rule which is worth mentioning regards the on site
value of the static structure factors, which is bound by the 
spin normalization condition:
\be
{1\over \b}\sum_{\om_n} \int {d^dk\over (2\pi)^d } \left [ F^{zz}(\bk,\om_n)
+ 2 F^{xx}(\bk,\om_n) \right ] + m^2 = S(S+1)
\label{norma}
\ee
at any temperature. 
If we choose to determine the parameters $\mu_\perp$ and $\mu_{||}$ 
by use of the susceptibility sum rules (\ref{stag}), in close analogy with
the procedure adopted for classical models \cite{hrt}, the additional
relation (\ref{norma}) cannot be generally satisfied by
the simple form (\ref{fij}). This deficiency is due to the presence of 
an incoherent part in the actual dynamic correlations which has been neglected
in our single mode approximation. 

A useful check on the accuracy of the parameterization (\ref{fij}) comes from the
zero temperature limit, where quantitative approximations are available,
like Spin Wave Theory (SWT). Although SWT is  formally an expansion around
the classical limit, even the simplest truncations in the 
parameter $1/S$ are known to be quite accurate even for $S=1/2$ 
\cite{mc2d}. To lowest order, the spontaneous staggered magnetization
attains its maximum value $m=S$ and
the correlation functions in imaginary
time can be easily calculated in vanishing external field $h$: 
Remarkably, the result has precisely the mean field structure (\ref{fij}) with 
$\mu_\perp=\mu_{||}=2Jd$, as expected in the symmetry broken state.
However, it is known that quantum 
fluctuations affect the zero field magnetization, the
spin wave spectrum $\epsilon_k$, the stiffness 
$\rho_s$ and the uniform susceptibility $\chi^\circ_\perp$. These 
quantities are indeed modified to first order in SWT \cite{igarashi},
although the form of $\epsilon_k$ (\ref{disp})
remains unaltered and only the prefactor is changed, leading to a 
simple renormalization of the spin wave velocity $c_s$:
\beqa
c_s=&Z_c\, JS\sqrt{4d}\,;  \hfill  &
Z_c=1+{1\over 2S}\int {d^dk\over(2\pi)^d } \ll [1-\e_{sw}(\bk)\rr ]
\nm\\
\rho_s=&Z_{\rho}\,JS^2\,;  \hfill &Z_\rho=
1-{1\over 2S}\int {d^dk\over(2\pi)^d } {\ll [1-\e_{sw}(\bk)\rr ]^2
\over \e_{sw}(\bk)}
\nm\\
\chi^\circ_\perp =& Z_\chi (J4d)^{-1}\,; \hfill &Z_\chi=
1-{1\over 2S}\int {d^dk\over(2\pi)^d } {1-\e_{sw}(\bk)^2
\over \e_{sw}(\bk)}
\label{swt}
\eeqa
where the momentum integrals are restricted to the lattice Brillouin zone
and 
\be
\e_{sw}(\bk)=\sqrt{1-d^{-2}\g_{\bk}^2}
\label{sweps}
\ee
A better representation of the dynamic correlations which, keeping the
single pole structure (\ref{singlem}), is able to reproduce the first order
SWT results (\ref{swt}) can be obtained by including two more adjustable 
renormalization amplitudes $\z$ in the functions $F^{ij}(\bk,\om_n)$:
\beqa
&& F^{xx} (\bk,\omega) =\z_\perp {\mu_\perp-\ww(\bk) \over  m^{-2}
\z_\perp^2\omega^2+ \mu_\perp^2-\ww(\bk)^2}   \nm  \\
&&  \nm \\
&& F^{xy}(\bk,\omega)=\z_\perp{\z_\perp m^{-1}\omega\over
m^{-2}\z_\perp^2\omega^2+\mu_\perp^2-\ww(\bk)^2}  \nm \\
&& \nm \\
&& F^{zz}(\bk,\omega)=\z_{||} {\delta_{\omega,0}\over\mu_{||}+\ww(\bk)}
\label{newfij}
\eeqa
\nn
At zero temperature, the full one loop SWT corrections
(\ref{swt}) are reproduced by the form (\ref{newfij}) with the choice
$\z_\perp=Z_\chi$  showing that such an ansatz for the 
dynamic correlation functions may improve the simple mean field description
of \HH antiferromagnets. However, in order to implement  
this generalization we need additional sum rules for determining 
self-consistently the parameters $\z_\perp$ and $\z_{||}$, 
besides the already mentioned Eqs. (\ref{stag}) for $\mu_\perp$ 
and $\mu_{||}$. A natural choice 
would be the analogous Eqs. (\ref{homo}), which control the
renormalization of the uniform susceptibility. However, we will 
not explicitly pursue this extension in the present work, but rather confine our
analysis to the simplest choice $\z_\perp=\z_{||}=1$, i.e. keeping the 
same structure of the dynamic correlations at the mean field
level (\ref{fij}), just allowing for ``mass renormalization", according
to Eqs. (\ref{stag}).
The single mode structure of the dynamic correlations,
besides reproducing the lowest order spin wave result at $T=0$,
has also been proven quite accurate in numerical studies on the 
\HH model both via series expansions \cite{singh} and Monte Carlo
simulations \cite{mc3d}. 

As noticed by Hertz in his pioneering work on phase transitions
in quantum systems \cite{hertz}, a proper treatment of critical 
phenomena in quantum mechanics directly provides predictions
on dynamic scaling in the system. The same feature is clearly
present in our microscopic approach which requires the form of
the dynamic structure factor. Unfortunately, in the approximation 
we are examining, 
the single pole form (\ref{singlem}) does not allow for 
anomalous scaling in the dispersion relation ($\om\sim k^z$) 
at the critical point. The functional dependence on 
$k$ and $\om$ in Eq. (\ref{disp}) correctly predicts a
linear spin wave dispersion $\om\sim c_s k$ in the symmetry broken 
state at all temperatures $T<T_c$. On approaching $T_c$, however, the 
spin wave spectrum vanishes as the order parameter $\e_k\sim m \sim
t^\b$ where $t=(T_c-T)/T_c$ and $\b$ is the appropriate critical exponent.
Therefore, exactly at $T_c$, the spin wave spectrum fades away and above $T_c$ 
only a delta contribution at $\om=0$ survives in the paramagnetic dynamic
correlations. Such a description
of dynamic critical behavior provided by the approximate form 
(\ref{singlem}) is clearly rather schematic and unsatisfactory.
However, the way the spin velocity vanishes $c_s\sim t^\beta$ 
is consistent with the predictions of dynamic scaling 
$c_s\sim \xi^{1-z}$ if the dynamic critical exponent is given by $z=d/2$. 
In fact, by inserting the power law behavior of the correlation
length $\xi\sim t^{-\nu}$ near the critical point,
we obtain $c_s\sim t^{\nu(d-2)/2}$, which 
gives $c_s\sim t^\beta$ via standard hyperscaling relations. The latter
equality holds because in our approach the correlation exponent $\eta=0$
due to the analytic $k$ dependence of the correlation functions. 
Remarkably, the result $z=d/2$ agrees with the 
general expectations of dynamic scaling for antiferromagnets: 
The appropriate equation of motion corresponds to
``model G" in which the order parameter does not commute with 
the hamiltonian but a conserved quantity (i.e. the uniform 
magnetization) is also present \cite{hohenberg}.

The discussion presented in this Section
supports the choice of a simple single mode form as a reasonable
approximation of the actual dynamic structure factor in \HH
antiferromagnets and leads us to propose Eqs. (\ref{fij},\ref{stag}) as
a closure to the exact renormalization group equation (\ref{qhrt}),
closely following the strategy successfully adopted in classical models
\cite{hrt}. In fact, the thermodynamic relationship (\ref{stag}) between
correlations and derivatives of the free energy holds independently
of the form of the interaction $\ww(\bk)$ and then it is valid for every 
$Q$-system, which corresponds to a potential $\ww_Q(\bk)$ characterized by a 
cut-off $Q$ on its Fourier components. Therefore, the dynamic correlation
functions of the $Q$-systems $F_Q^{ij}$ can be parametrized by the same form
(\ref{fij}) used at $Q=d$ with  $Q$-dependent parameters 
$\mu_{\perp Q}$ and $\mu_{|| Q}$ defined by the analog of Eqs. (\ref{stag}):
\beqa
\left({h_Q\over m}\right)^{-1}&=&
m\left ({\partial a_Q\over\partial m}\right )^{-1} =
F_Q^{xx}(\bg,0)=(\mu_{\perp Q} - 2Jd)^{-1} \nm\\
\left ({\partial h_Q\over \partial m}\right)^{-1}&=&
\left ( {\partial^2 a_Q\over \partial m^2 }\right)^{-1}=
F_Q^{zz}(\bg,0)=(\mu_{|| Q} - 2Jd)^{-1} 
\label{stagq}
\eeqa
When the {\sl approximate} form (\ref{fij}, \ref{stagq})
of the dynamic correlations is inserted in the 
{\sl exact} evolution equation (\ref{qhrt}) we get a {\sl closed} equation
for the evolution of the free energy of the \HH model. The
frequency sums in Eq. (\ref{qhrt}) can be analytically evaluated by
Euler summation formula while the momentum integration can be trivially
carried out with the result:
\be
\frac{d \, a_Q}{dQ}=\frac{D_d(Q) }{2\beta}
\left \{ 4\,\ln \left [{\sinh\left ({1\over 2}\beta m\mu_{\perp Q}\right )\over
\sinh\left ({1\over 2}\beta m\sqrt{\mu_{\perp Q}^2-4J^2(d^2-Q^2)
}\right )}\right ]
+ \ln\left [{\mu_{|| Q}^2\over\mu_{|| Q}^2-4J^2(d^2-Q^2)} \right]\right \}
\label{qhrt2}
\ee
where the ``density of states" $D_d(Q)$ is defined as
\be
D_d(Q)={1\over 2}\int {d^d k\over (2\pi)^d}\,\, \d(\sqrt{d^2-\g(\bk)^2}-Q)
\label{dos}
\ee
This evolution equation will be analyzed in the following Sections. 

Finally, it is interesting to consider the limiting form of
our equation (\ref{qhrt2}) when $S\to\infty$, i.e. when the quantum 
\HH model reduces to a classical $O(3)$ vector model. Clearly,
we have to rescale all the dimensional quantities by suitable powers
of $S$ in order to get a finite result. In particular, the
magnetization $m\to Sm$ while energy and temperature must be scaled as
$a\to S^2 a$ and $\b\to S^{-2}\b$. As a consequence, equation
(\ref{qhrt2}) simplifies and we get
\be
\frac{d \, a_Q}{dQ}=\frac{D_d(Q) }{2\beta}
\left \{ 2\,\ln \left [{\mu_{\perp Q}^2 \over
\mu_{\perp Q}^2-4J^2(d^2-Q^2)}\right ]
+ \ln\left [{\mu_{|| Q}^2\over\mu_{|| Q}^2-4J^2(d^2-Q^2)} \right]\right \} 
\label{qhrtc} 
\ee 
which in fact coincides with the equation obtained in the framework
of HRT for a classical vector model \cite{hrt}.

\section{Long wavelength analysis}\label{5}

Let us study the behavior of the QHRT evolution equation (\ref{qhrt2}) 
in the critical region. This investigation is particularly instructive
because it shows how universality and renormalization group (RG)
come about in the QHRT framework. We will consider the
critical region of the paramagnetic-antiferromagnetic transition,
where the physical staggered susceptibilities ($\chi^s_\perp$, $\chi^s_{||}$) 
diverge. First we remark that our ansatz for the two point 
function (\ref{fij}) is analytic in $\bk$. As a consequence, the
correlation exponent predicted by our approach is $\eta=0$ in
any dimension. It is known that such a result is correct 
only above the upper critical dimension $d=4$. However, due to the
smallness of $\eta$ in $d=3$ ($\eta\sim 0.04$ \cite{zinn}), 
we consider such an approximation acceptable also in the physically
relevant applications.

In order to study the critical behavior predicted by our approach,
we now focus our attention on the last stages of the evolution 
($Q\sim 0$), i.e. on the long wavelength form of equation (\ref{qhrt2}).
In fact, we know that the order parameter fluctuations of wavevector $\bk$
close to $\bg$ are inhibited by the presence of a cut-off $Q$ and
the free energy of a $Q$-system cannot develop singularities 
at finite $Q$, but only 
in the $Q\to 0$ limit. Eqs. (\ref{stagq}) show that, in this
regime, both $\mu_{\perp Q}$ and $\mu_{|| Q}$ tend to
$2dJ$. Moreover, being the critical point at $m=0$, 
our analysis will be limited to the $m\sim 0$ region.
The asymptotic form of the QHRT equation (\ref{qhrt2}) is then:
\beqa 
\frac{da_Q}{dQ}=  -\frac{\Omega_d }{2\b(2\pi\sqrt{d})^d}
Q^{d-1} \ll\{2 
\ln \ll [ m^{-1}{\p a_Q \over \p m }+J{Q^2\over d}\rr ]+
\ln \ll [ {\p^2 a_Q \over \p m^2 }+J{Q^2\over d}\rr ]
\rr \} 
\label{rg0} 
\eeqa
where $\Omega_d$ is the surface of the unit $d$-dimensional hypersphere.
Remarkably, this limiting form of the evolution equation coincides with the 
one appropriate for a classical $O(3)$ vector model in $d$ dimension, 
i.e. with the long wavelength limit of Eq. (\ref{qhrtc}). This 
confirms that Quantum Mechanics becomes irrelevant in the neighborhood 
of a (finite temperature) critical point \cite{hertz}. 

In order to exploit the renormalization group structure hidden in Eq.
(\ref{rg0}) we suitably rescale the variables:
\beqa
z&=&\ll[ \frac{d\,\Omega_d}{2\b J (2\pi\sqrt{d})^d} \rr] ^{-1/2}
Q^{-(d-2)/2}  m \nm\\
H&=&\ll[\frac{\Omega_d}{2\b(2\pi\sqrt{d})^d}\rr] ^{-1} Q^{-d} 
[a_Q(m)-a_Q(m=0)]  
\label{rescale} 
\eeqa
In this way, the explicit dependence of the equation on $Q$ is
eliminated and the evolution equation becomes:
\be 
\dot H =dH+z\frac{2-d}{2} H' +   \ll [ 
\ln \frac {H''+1}{H''(0)+1} +2 \ln \frac{H'/z+1}{H''(0)+1} \rr ]
\label{rg1} 
\ee 
where dots represent derivatives with respect to $l=\ln(d/Q)$ 
and primes derivatives in $z$.
This equation is formally identical to that obtained 
by Nicoll {\it et al.} in an approximate renormalization
group formulation \cite{nicoll} and, as such, it 
has already been studied by means of standard RG techniques. 
This analysis showed that the critical point 
is present only for $d>2$, in agreement with Mermin-Wagner theorem
\cite{mermin}. Note that precisely at $T=0$, the QHRT equation (\ref{qhrt2})
does not reduce to the asymptotic form (\ref{rg0}) and therefore
quantum critical points would require a different long wavelength
analysis. Eq. (\ref{rg1})
properly embodies critical scaling and hyperscaling and,
as shown in Ref. \cite{nicoll}, it
correctly gives non classical critical exponents
to first order in the $\e=4-d$ expansion. For instance, the
staggered susceptibility exponent at $m=0$ ($\chi^s_{||}\sim t^{-\g}$) is:
\be
\g = 1+ \frac{5}{22} \e +O(\e^2)
\ee
The value of $\g$ predicted by our approach in $d=3$ can be found
by integrating the fixed point equation associated to 
(\ref{rg1}) and solving the corresponding eigenvalue problem.
The numerical result gives $\g=1.65$, to be compared with the
accepted value for a three component order parameter $\g=1.39$ 
\cite{zinn,landau}. The other critical exponents can be obtained from $\g$ 
by scaling, recalling that $\eta=0$ in our approximation. 

Interestingly, a similar RG analysis can be carried out on the same
asymptotic QHRT equation (\ref{rg0}) in order to study the 
divergence of the susceptibility on approaching the {\it coexistence}
curve at $T<T_c$. By rescaling $m$ in the neighborhood of the
spontaneous magnetization $m_{\times}$ as
\beqa
z&=&\ll[ \frac{d\,\Omega_d}{2\b J m_\times (2\pi\sqrt{d})^d} \rr] ^{-1}
Q^{-(d-2)}  (m-m_{\times}) \nm\\
H&=&\ll[\frac{\Omega_d}{2\b(2\pi\sqrt{d})^d}\rr] ^{-1} Q^{-d} 
[a_Q(m)-a_Q(m_{\times})]  
\label{rescalex} 
\eeqa
the evolution equations at long wavelengths reads:
\be 
\dot H =dH+z(2-d) H' +   \ll [ 
\ln \frac {H''}{H''(0)} +2 \ln \frac{H'+1}{H'(0)+1} \rr ]
\label{rgx} 
\ee 
At coexistence, the free energy approaches a fixed point 
which corresponds to a stationary solution of 
of Eq. (\ref{rgx}), regular in the whole interval 
$z\in (0,\infty)$ and satisfying the boundary conditions
$H(0)=0$ and $H(z)\to z^{d/(d-2)}$ as $z\to +\infty$.
The latter requirement is necessary in order to match with
the solution outside the coexistence region (i.e. at $m>m_{\times}$).
As a consequence, the staggered susceptibilities display a 
non-analytic behavior when $d<4$:
\be
\chi_\perp^s\sim (m-m_\times)^{-2/(d-2)} \,;\qquad 
\chi_{||}^s\sim (m-m_\times)^{-(4-d)/(d-2)}
\label{coex}
\ee
in agreement with the predictions of a spin wave analysis
in the classical $O(3)$ model \cite{zinn}.
 
Let us conclude this section showing how it is possible to recover, 
within the QHRT formalism, also the RG equations for the coupling 
constants of the quantum non linear sigma model ($QNL\sigma M$).   
The long-wavelength behavior of the $d$ dimensional 
\HH antiferromagnet in zero external field
is commonly believed to be described by a $d+1$ 
dimensional $QNL\sigma M$, at least in the low temperature regime
\cite{haldane}. This is an effective field theory in which
the additional temporal variable, defined only in the 
interval $(0,\b)$, reflects the presence of quantum fluctuations. 
The $QNL\sigma M$ action is conventionally written in terms of
a field ${\bf \Omega }(\br,t)$ as:
\be
S[{\bf \Omega }]={\rho_s\over 2}\int d^d \br \int_0^\b dt 
\,\left\{\vert \nabla {\bf \Omega }\vert^2
+{1\over c_s^2} \left ({\p {\bf \Omega }\over \p t}\right)^2\right\}
\label{nlsm}
\ee
where the local constraint $\vert {\bf \Omega }\vert =1$ is understood
and $\rho_s$ and $c_s$ are the bare (i.e. mean field) spin stiffness and 
velocity. The phase diagram of the model depends on temperature and on the
dimensionless coupling constant:
\be 
g = \frac{c_s \L ^{d-1}}{ \r_s} 
\label{gzero} 
\ee
where $\L$ is the ultraviolet momentum cut-off, which is of the
order of the reciprocal of the lattice spacing. The $QNL\sigma M$,
as a low energy theory for \HH antiferromagnets, has been extensively 
studied by Chakravarty Halperin and Nelson (CHN) \cite{chn} by 
weak coupling RG analysis. 
The one loop  equations describing the 
evolution of the couplings constants are: 
\beqa 
&& \frac{dg}{dl}=(1-d)g+\frac{K_d}{2}g^2 \coth(g/2t) \nm \\
&&   \frac{d}{dl}\ll ( \frac{g}{t}\rr)= -\frac{g}{t} 
\label{wrg}  
\eeqa
\nn 
Here, $e^l$ is the length rescaling factor and $K_d$ is a geometrical 
coefficient 
($K_d^{-1}=2^{d-1}\pi^{d/2} \Gamma (d/2)$). 
The equations are written for the running coupling constant $g(l)$
and for the dimensionless temperature scale $t(l)$ whose initial condition is
$t=\L^{d-2}/(\b \r_s)$.
For $d\leq 2$ there are no finite temperature fixed points, while
at $T=0$ and for $d>1$ there is a non trivial fixed point
given by $g_c=2/K_d(d-1)$ describing a quantum transition.  
Depending on the initial value of the coupling constant $g$,
the RG flow drives $g$ either to zero or to infinity. In the former case
the physics is governed by the low temperature fixed point and then 
the system is ordered. Assuming that the two dimensional \HH model 
has a symmetry breaking ground state, from the solution
of the RG equations (\ref{wrg}), the following behavior of 
the correlation length is obtained \cite{chn}:
\be  
\xi(T)  \sim \xi_0 {\rm e}^{\frac{2\pi \bar\r_s}{T}} 
\label{xiditi} 
\ee
where $\bar\r_s=\rho_s(g_c-g)/g_c$ 
is the renormalized spin-stiffness at $T=0$.

The weak coupling RG analysis predicts
several regimes where quantum and thermal fluctuations 
play different roles. However, quantitative estimates 
based on low energy effective theories are usually rather inaccurate
because short wavelength fluctuations are neglected in these approaches.
Here we want to show that the QHRT equation, 
reduces, at low energy, precisely to the one loop RG equations studied by CHN
(\ref{wrg}). However, QHRT provides a description of spin fluctuations over 
{\it all} lengthscales maintaining the advantages of a fully microscopic theory.

\begin{figure}[htbp]
\protect 
\centerline{\psfig{figure=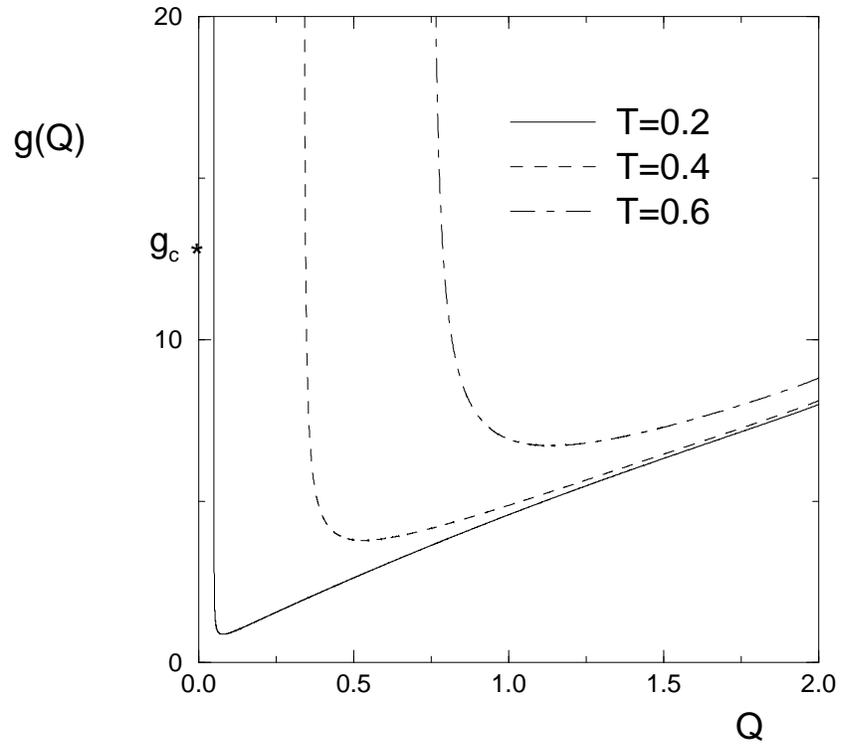,width=11cm}} 
\caption{\it
RG trajectories for the two dimensional Heisenberg model
computed via numerical integration of the QHRT equation.}
\protect
\label{fig1}
\end{figure}

In the low temperature region, longitudinal fluctuations can
be neglected because they are characterized by a less singular behavior and
therefore, in this regime, the last term in Eq. (\ref{qhrt2}) may be
dropped. If we now differentiate Eq. (\ref{qhrt2}) 
with respect to the magnetization, we get an equation for the 
evolution of the staggered field $h_Q(m)$ at fixed $m$: 
\beqa
\frac{dh_Q}{dQ}&=&D_d(Q) \Bigg [
\coth \ll (\frac{\b}{2} m \mu_{\perp Q}\rr ) \mu_{|| Q} \nm\\
 &-& \coth \ll (\frac{\b m}{2} \sqrt{\mu^2_{\perp Q} -4J^2(d^2-Q^2)} \rr)
\frac{\mu _{\perp Q}\mu_{|| Q} -4J^2(d^2-Q^2)}
{\sqrt{\mu^2_{\perp Q}-4J^2(d^2-Q^2)}} \Bigg ]
\label{wqhrt}
\eeqa
This equation implicitly describes 
how the spontaneous magnetization is reduced from
its mean field estimate when fluctuation are included. In order to 
exploit this information, it is 
useful to define a $Q$-dependent spontaneous magnetization $m_Q$ by 
requiring that $h_Q(m_Q)=0$. Such a quantity satisfies an evolution
equation which can be easily deduced from Eq. (\ref{wqhrt}).
At long wavelengths (i.e. keeping the leading terms as $Q\to 0$) 
the $Q$ dependent staggered magnetization in zero field obeys a 
closed equation:
\be 
\frac{d m_Q}{d Q} = K_d \ll(\frac{Q}{\sqrt{d}}\rr) ^{d-2} \coth(\b Q m_Q )
\label{mag} 
\ee
In order to make contact with the RG picture based on the $QNL\sigma M$
effective action, we now have to relate the QHRT variable $m_Q$ 
with the running coupling constant $g(l)$ appearing in Eqs. (\ref{wrg}).
The correspondence between the length rescaling factors is clearly
$l=\ln(d/Q)$, while
the definition of the coupling constant $g$ in terms of physical 
quantities, like stiffness and spin velocity, given in Eq. (\ref{gzero})
suggests the proportionality between $g^{-1}$ and the spontaneous
magnetization. In fact, as discussed in Section (\ref{4}), QHRT
provides explicit relations between $\r_s$, $c_s$ and the 
magnetization at coexistence. It is therefore rather natural to introduce
an ``effective coupling constant" $g_Q=\sqrt{4d}(Q/\sqrt{d})^{d-1}m_Q^{-1}$
which, when substituted in Eq. (\ref{mag}), satisfies an equation
formally identical (to order $g^2$) with the standard one loop RG
form (\ref{wrg}). 
As an example, we plot in Figure [\ref{fig1}] the RG flux of $g_Q$ as
obtained by the integration of the {\sl full} QHRT equation in $d=2$
(\ref{qhrt2}) via the previous definition of $g_Q$.
We clearly see the effect of the unstable zero temperature weak
coupling fixed point while, for the nearest neighbor Heisenberg model,
the other fixed point ($g_c=4\pi$), governing the Quantum Critical
regime, has no effect on the RG trajectories. This suggests that 
the \HH model always remains in the Renormalized Classical regime
in agreement with an independent analysis based on the 
pure quantum self consistent harmonic approximation (PQSCHA) \cite{tognetti}.

\section{Numerical results} \label{6}
\nn
Equation (\ref{qhrt2}), together with the thermodynamically consistent closure
relations (\ref{stagq}) and the appropriate initial condition at $Q=d$
(\ref{inizio}) has been solved numerically in two and three dimensions
and several values of the spin $S$. Following the analogous treatment of
HRT in classical systems \cite{hrt} we first wrote the partial
differential equation in quasilinear form and then used a fully
implicit finite difference scheme \cite{ames}.
A careful implementation of the
numerical algorithm is necessary, because the solution to the QHRT
equation develops singularities as $Q\to 0$, when a phase boundary is crossed.
This is a consequence of the special treatment of long wavelength 
fluctuations provided by QHRT which is able to reproduce rigorously flat 
isotherms inside the coexistence region, as already shown in classical
systems \cite{pini}. Our computations were carried out with a
mesh of few hundred points along the magnetization axis, which 
already allows for a good accuracy.
The high temperature asymptotic behavior of thermodynamic quantities
is correctly reproduced by QHRT in arbitrary dimension $d$ and for
every spin $S$, as can be easily shown by analysis
of the evolution equation (\ref{qhrt2}).
Before discussing the numerical results in three dimensions
we briefly comment on the calculations performed on the square lattice.
In this Section we set $J=1$.

\subsection{The two dimensional case}

\begin{figure}[htbp]
\protect 
\centerline{\psfig{figure=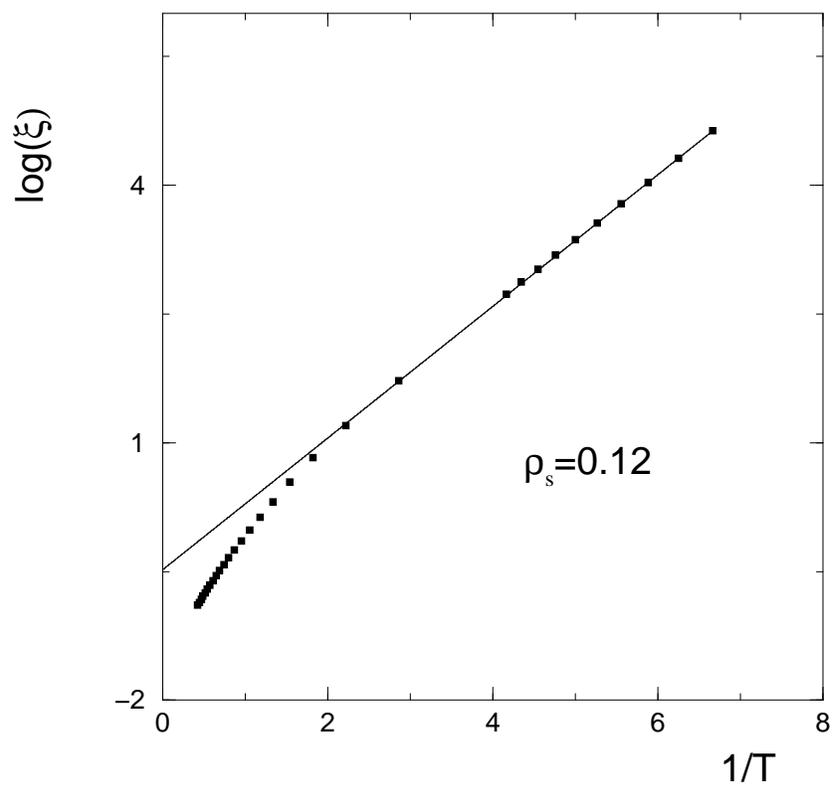,width=11cm}} 
\nn \caption{\it
Correlation length in two dimensions for $S=1/2$.}
\protect
\label{fig2}
\end{figure}

As already noticed, our equation correctly predicts the absence of
spontaneous symmetry breaking at finite temperature for every choice of the
spin $S$. However, the effective coupling constant flow shown in
Fig. [\ref{fig1}]
illustrates the strong attractive nature of the unstable zero temperature 
fixed point and allows to give a precise estimate of the ground state 
spontaneous magnetization. For $S=1/2$ we obtain  $m_{\times}\sim 0.35$,
rather close to the value $m\sim 0.307$ obtained by numerical simulations 
\cite{mc2d}. 
Weak coupling RG analysis predicts that the correlation length at $m=0$
diverges exponentially on approaching zero temperature
with a coefficient related to the spin stiffness (\ref{xiditi}). The same 
behavior is found by our numerical results as shown in Fig. [\ref{fig2}]. This 
result is not unexpected because we have already shown that QHRT contains
all the information provided by the one loop RG approach. The estimate
of the spin stiffness coming from a linear fit of the data in Fig. [\ref{fig2}]
gives $\rho_s\sim 0.12$ for $S=1/2$,
to be compared with the accepted value $\rho_s\sim
0.18$ \cite{mc2d}. Our result is in fact consistent with the relationship
$\rho_s=m_{\times}^2$ discussed in Section (\ref{4}) which follows from
the adopted closure. Therefore we are led to conclude that the discrepancy
between the QHRT and Monte Carlo (MC) in the estimated values of $\rho_s$
is probably due to the crude approximation for
the two point functions we have adopted in this calculation, as anticipated
in Section (\ref{4}), see also Eq. (\ref{newfij}).

\begin{figure}[htbp]
\protect 
\centerline{\psfig{figure=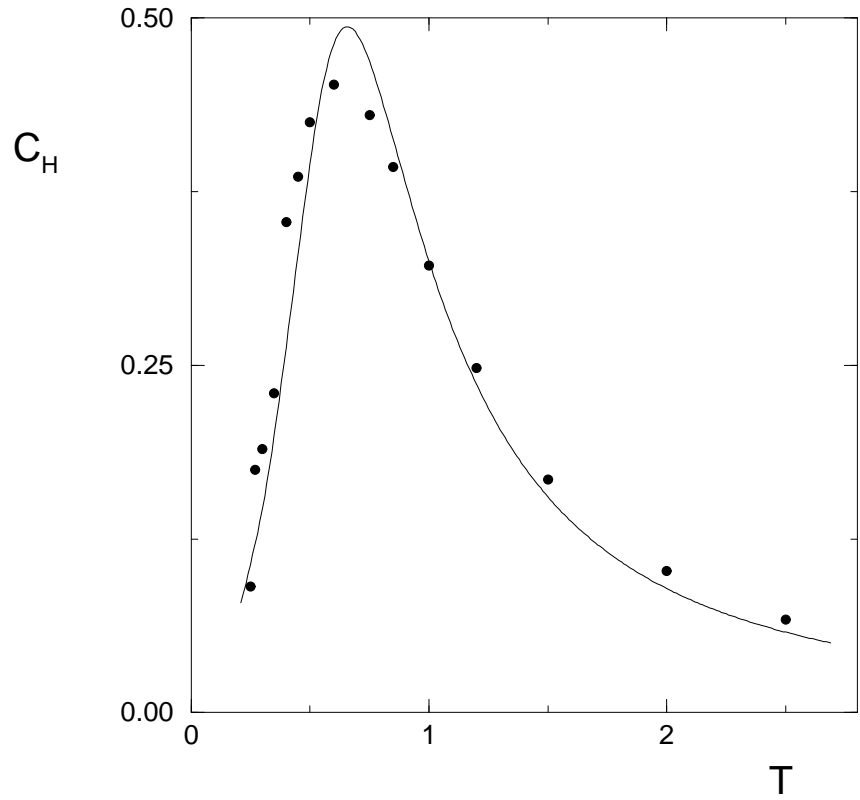,width=11cm}} 
\caption{\it
Zero field specific heat $C_H$ as a function of temperature
for the $S=1/2$ case in two dimensions.
Full line QHRT, circles Monte Carlo simulations \protect \cite{makivik}.}
\protect
\label{fig3}
\end{figure}

The zero field specific heat $C_H$ as a function of temperature is shown in
Fig. [\ref{fig3}] together with a comparison with MC data for $S=1/2$ 
\cite{makivik}. 
The non monotonic behavior, even in the absence of a
phase transition, is just a consequence of the 
asymptotic decay of $C_H$ at high temperature ($C_H\propto T^{-2}$)
and to the vanishing of $C_H$ at $T=0$, as required by thermodynamics.

\subsection{The \HH model on a cubic lattice}

\begin{figure}[htbp]
\protect 
\centerline{\psfig{figure=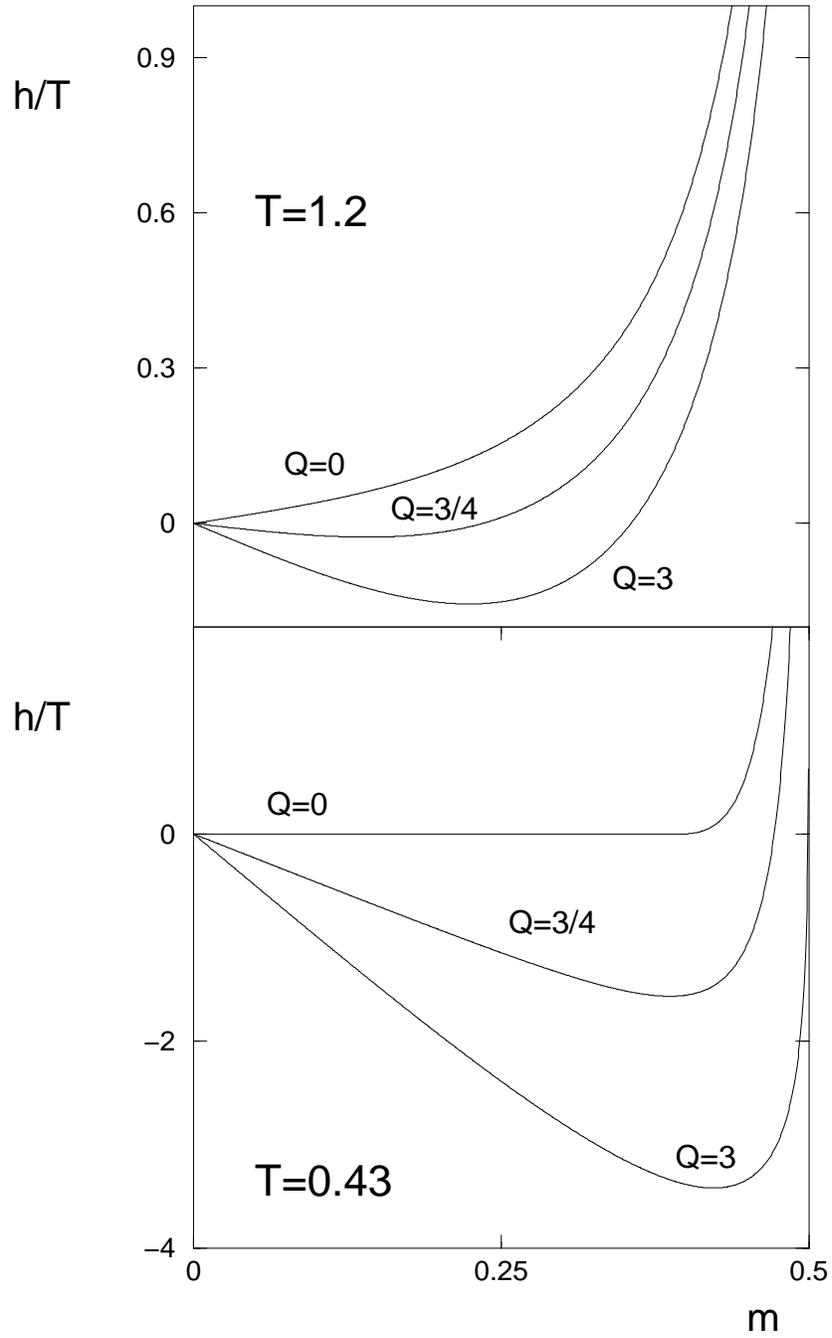,width=11cm}} 
\caption{\it Behavior of the equation of state $h(m)$ at two different
temperatures as a function of the evolution parameter $Q$. 
Above $T_c\,<\,T=1.2 \, < T_{\rm MF}=1.5$, being   $T_{\rm  MF}$   
the critical temperature in Mean-Field approximation.
Below $T=0.43\,<T_c$.}
\protect
\label{fig4}
\end{figure}

In $d=3$, spontaneous symmetry breaking at finite temperature 
does occur in the \HH model. As a consequence, thermodynamics
forces the Helmholtz free energy to be rigorously flat inside the
coexistence region. This feature is usually lost in the approximate
treatments of the model and may be recovered {\it a posteriori} 
by Maxwell double tangent construction. As already noticed in
the application of HRT to classical statistical models, our 
approach instead is able to predict rigorously flat isotherms 
in a region of the phase diagram, thereby allowing for a 
unambiguous determination of the coexistence curve \cite{pini}. 
As an example, in Fig. [\ref{fig4}] we show the behavior 
of the equation of state
$h(m)$ at two temperatures: above and below $T_c$
for several values of the evolution parameter $Q$. We see that, although
the initial condition at $Q=3$ is in both cases below the mean field
critical temperature (i.e. corresponding 
to a negative mean field transverse susceptibility), $\chi_{\perp}^{-1}=h/m$ 
is strongly renormalized by fluctuations and for $T>T_c$ it
becomes positive and saturates as $Q\to 0$, while for $T<T_c$ eventually 
sticks to zero. Therefore, during the evolution, the region characterized by
negative susceptibility is gradually removed (for $T> T_c$) or replaced
by a flat isotherm (for $T< T_c$).
This behavior can be analytically understood from a study of the
asymptotic equation, in close analogy to the Ising case \cite{pini}. 

\begin{figure}[htbp]
\protect 
\centerline{\psfig{figure=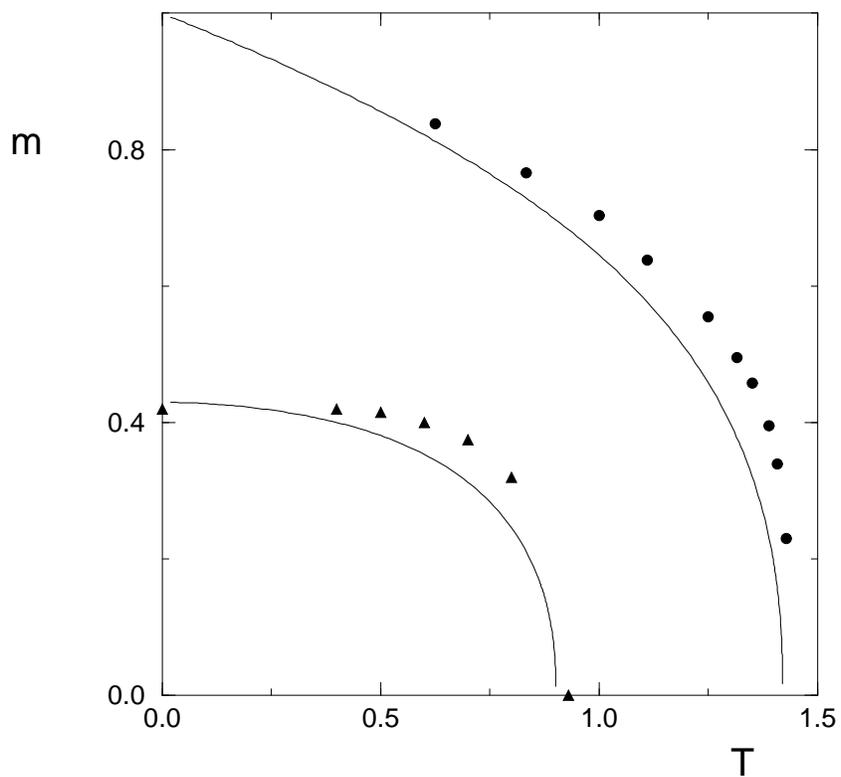,width=11cm}} 
\caption{\it  
Above: Coexistence curve of the classical \HH model. Full line QHRT, circles
Monte Carlo simulations \protect \cite{binder}. 
Below: Coexistence curve of the $S=1/2$ \HH model. Full line QHRT, triangles
series expansion \protect \cite{kok}.}
\protect
\label{fig5}
\end{figure}

\begin{figure}[htbp]
\protect 
\centerline{\psfig{figure=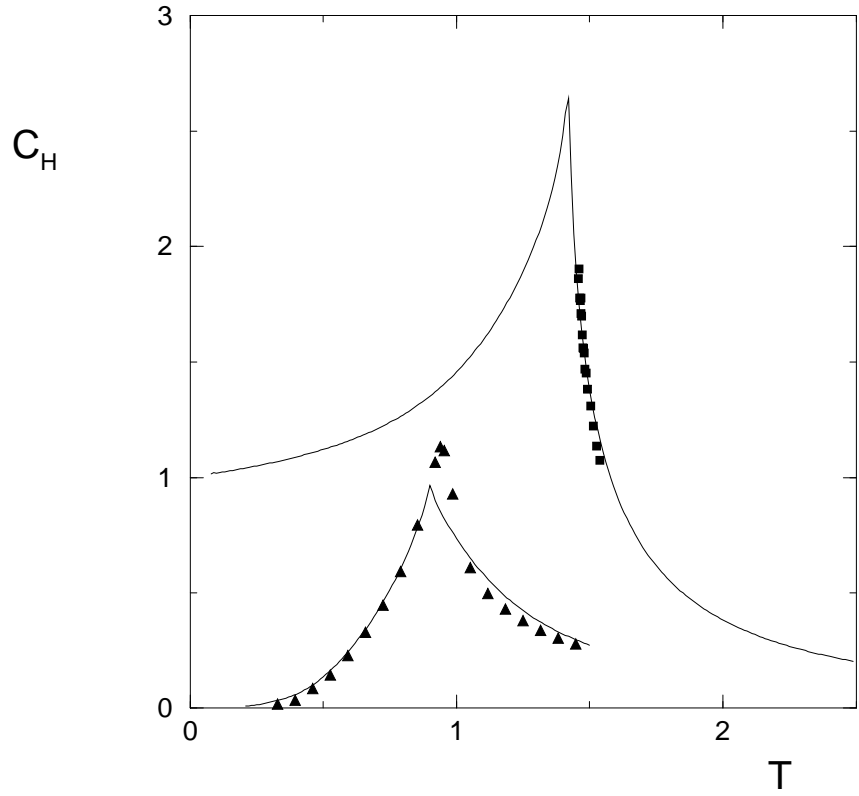,width=11cm}} 
\caption{\it  
Above: Specific heat of the classical \HH model. Full line QHRT,
squares Monte Carlo simulation   \protect \cite{holm}.
Below: Specific heat of the $S=1/2$ \HH model. Full line QHRT,
triangles Monte Carlo simulations \protect \cite{mc3d}.}
\protect
\label{fig6}
\end{figure}

\begin{figure}[htbp]
\protect 
\centerline{\psfig{figure=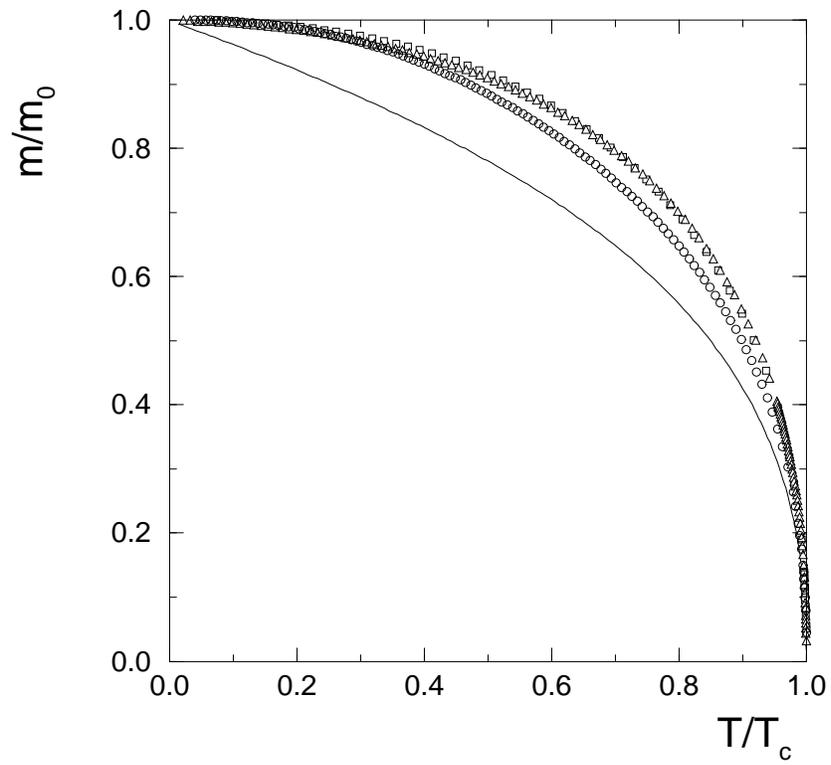,width=11cm}} 
\caption{\it  
Reduced spontaneous magnetization as a function of temperature
for different values of the spin: $S=1/2$ (triangles) $S=1$ (squares)
$S=5/2$ (circles) and $S=\infty$ (full line).}
\protect
\label{fig7}
\end{figure}

The full coexistence curve for $S=1/2$ and $S=\infty$ 
is reported in Fig. [\ref{fig5}] together
with the results of series expansions \cite{kok} and numerical
simulations \cite{binder}.
The zero temperature limit of the spontaneous magnetization 
gives $m_{\times}=0.43$ in our approximation, 
to be compared with the value $0.42$ from second order spin-waves theory
\cite{oguchi}. The predicted critical temperature is shown
in Table I for several values of the spin $S$. 
When available, also estimates from Monte Carlo, series expansions
or cumulant expansions are given.
In Fig. [\ref{fig6}] the specific heat as a function of the 
temperature is plotted for $S=1/2$ and $S=\infty$ 
together with classical MC simulations and recent QMC data \cite{mc3d}. 
Note that QHRT correctly reproduces the different zero temperature 
limit of the quantum ($C_H\to 0$) and classical ($C_H\to NJ$) systems.
Again, the agreement is quite satisfactory, being comparable with
simulation uncertainties. We remark that QHRT correctly predicts a
cusp-like behavior at $T_c$, although the corresponding critical
exponent $\alpha <0$ is underestimated by our approximation.
These small differences, however, cannot be appreciated on the
scale of the plot. 
Finally, Fig. [\ref{fig7}] shows the change in the shape of the 
coexistence curve due to the effects of quantum
fluctuations: going from the classical ($S=\infty$) to
the quantum $S=1/2$) case, the magnetization curve rounds off and the 
linear behavior at low temperature disappears, in agreement with 
SWT which predicts a quadratic temperature dependence of 
$m_{\times}$ as $T\to 0$.

In summary, we believe that even within the crude simple pole 
approximation adopted in the present study, QHRT is able to get
quantitative agreement with recent numerical data on the 
thermodynamics of quantum antiferromagnets in $d=3$. The error can be
estimated in few percents on the critical temperature, spontaneous
magnetization and specific heat. Larger discrepancies can be 
seen in the shape of the coexistence curve: such a disagreement is
rather unexpected and 
more accurate simulation data are needed before reaching a definite 
conclusion on this delicate issue. 
\section{Conclusions} \label{conclusions}

In this paper
we have introduced a new theoretical tool for  the study
of phase transitions in quantum systems.
Such an approach is based on the use of the Hierarchical Reference Theory 
(HRT) \cite{hrt}, a theory
developed in the context of classical fluids
able to correctly describe the
system in a wide region of the phase diagram, both
near and far from the critical point.
Fluctuations over different lengthscales are gradually
included thereby reproducing the full structure of Wilson
renormalization group theory in the critical region.
The possibility of applying HRT also in the context of quantum
systems (QHRT) follows from the observation that
the grand partition function of a wide class of quantum models
can  be formally mapped onto a classical partition
function (in $d+1$ dimensions) to which HRT can be directly applied.
The advantage of our formulation is that
since this mapping is exact, the microscopic
informations are correctly included into this formalism
and only in the vicinity of the critical
point and at long wavelengths the peculiar features of 
the model disappear and the evolution equations acquire
a universal form. 
QHRT is a general approach which can in principle be applied to a
large class of systems including fermions, bosons and spins.

The validity of this theory has been studied 
in the specific case of the isotropic \HH
antiferromagnet. This choice is dictated by the large amount of
available results from other theories and numerical methods, which
include mean field theories \cite{manousakis}, 
renormalization group \cite{chn} and path integral approaches \cite{tognetti},
series expansions \cite{kok} or Quantum Monte Carlo simulations
\cite{mc3d}.
In order to obtain a closed and tractable equation out of the 
QHRT formalism,
we analyzed a simple approximation which amounts to parameterize the
dynamic correlations in a single pole form. Even within this
approximation scaling and hyperscaling are satisfied 
close to the critical point, the critical exponents
are correct near four dimensions and, in $d=3$, take 
non classical values slightly larger than the accepted ones. 
Remarkably, the weak coupling renormalization group equations
for the Non Linear Sigma Model are recovered at low temperature
even in $d=2$, thereby reproducing the correct asymptotic
scaling as $T\to 0$.  

The theory has been tested against numerical data 
on a simple cubic lattice with nearest neighbors interaction  
and for several values of the spin.
The results are in close agreement with available simulations for
several quantities like the location of the critical
point, the spontaneous magnetization and the specific heat. 
In  $d=2$ the comparison with simulations is less satisfactory
even if QHRT is able to reproduce the qualitative
behavior of the model.
The discrepancies we found are related to the single mode
approximation adopted in the closure of the equation 
and may be overcome by use of a more flexible form for
the dynamic structure factors. 

In this paper our attention has been focused on the 
\HH model since it allows for
several interesting generalizations, like the introduction of
anisotropies or competing interactions. In these cases,
quantum and thermal fluctuations may play a more fundamental
role in modifying the topology of the phase diagram. 
A detailed study of the \HH model with easy plane anisotropy
is under way: the results will shed light on the thermodynamics of
the crossover between two and three component order parameter,
with direct experimental implications.
An extremely interesting application of the QHRT formalism to 
fermionic systems regards the Hubbard model, whose hamiltonian 
has in fact the general structure of Eq. (\ref{hhh}). The natural choice,
as a reference system, is the free Fermi gas on the lattice in  an
external magnetic field, while the Hubbard interaction couples the 
on site electron densities. In this case, the evolution
equations will involve density and magnetization fluctuations
allowing for a characterization of the phase transitions which 
lead to the creation of magnetic structures (ferro, antiferro or
itinerant) or charge density wave instabilities (phase separation
or stripe formation). 
We believe that a theory able to deal with long wavelength 
fluctuations, like QHRT, may represent a useful tool 
for tackling this class of problems.

\clearpage
\vfill\eject

\vspace*{0.5cm}

\begin{table}
\begin{center}
\begin{tabular}{|c|c|c|c|}
\hline
& QHRT & Monte Carlo & Series \\
\hline
$S=1/2$ & $ 0.90$ &$0.94$ \cite{mc3d} & $0.93 $  \cite{kok}  \\
\hline
$S=1$ &$2.66$  & &   \\
\hline
$S=3/2$ &$5.16$  & &  \\
\hline
$S=2$ &$8.35$  & &  \\
\hline
$S=5/2$ &$12.25$ & &  \\
\hline
$S=\infty$ & $1.419$ &  $1.443$ \cite{landau} & $1.445$ \cite{domb}  \\
\hline
\end{tabular}
\vskip 0.5 true cm
          \caption{
\it {Critical temperatures for different values of the
spin obtained by QHRT and other methods in the 
three dimensional Heisenberg model.}}
\end{center}
\end{table}

\np
\vspace*{2.0cm}

\np
\vspace*{2.0cm}

\np
\vspace*{2.0cm}

\np
\vspace*{1.0cm}

\np
\vspace*{2.0cm}

\np
\vspace*{2.0cm}

\np
\vspace*{2.0cm}
\end{document}